\documentstyle[aps,twocolumn]{revtex}
\begin{document}
\twocolumn[\hsize\textwidth\columnwidth\hsize\csname@twocolumnfalse\endcsname

\title
{Quantum nonlocality and applications in quantum information processing
of hybrid entangled states}

\author{Zeng-Bing Chen,$^1$ Guang Hou,$^1$ and Yong-De Zhang$^{2,1}$}
\address
{$^1$Department of Modern Physics, University of Science and Technology of China,
Hefei, Anhui 230027, P.R. China}
\address
{$^2$CCAST (World Laboratory), P.O. Box 8730, Beijing 100080, P.R. China}
\date{Received 12 March 2001}
\maketitle 

\begin{abstract}

\
The hybrid entangled states generated, e.g., in a trapped-ion or atom-cavity system, 
have exactly one ebit of entanglement, but are not maximally entangled. 
We demonstrate this by showing that they violate, but in general do not maximally violate,
Bell's inequality due to Clauser, Horne, Shimony and Holt. These states are 
interesting in that they exhibit the entanglement between two distinct degrees of 
freedom (one is discrete and another is continuous).
We then demonstrate these entangled states
as a valuable resource in quantum information processing including quantum 
teleportation, entanglement swapping and quantum computation with ``parity 
qubits". Our work establishes an interesting link between quantum information 
protocols of discrete and continuous variables.

\


\
\end{abstract}]

In the burgeoning field of quantum information theory \cite{QIT,Nature},
many practical applications heavily depend on quantum entanglement \cite
{EPR,Schro} as a necessary resource. Initially most of the concepts (e.g.,
quantum teleportation \cite{telep-d,Italy}, quantum error correction \cite
{error-d}, entanglement swapping \cite{telep-d,swap-d} and quantum
computation \cite{computation-d}) in quantum information theory were
developed for qubit systems with discrete quantum variables. Quantum
information protocols (including quantum teleportation \cite{BK}, quantum
error correction \cite{error}, quantum computation \cite{computation} and
entanglement swapping \cite{swap}) for continuous quantum variables have
also been proposed very recently in parallel.

Another issue closely related to quantum entanglement is quantum
nonlocality. Starting from local realism, Einstein, Podolsky and Rosen (EPR)
argued the incompleteness of quantum mechanics. Based on Bohm's \cite{Bohm}
version of the EPR entangled states Bell derived his famous inequalities 
\cite{Bell,CHSH,Bell-book}, enabling to test quantum mechanics against local
reality \cite{Aspect}. However, further studies of quantum nonlocality used
mainly Bohm's version \cite{Bohm} of the EPR states instead of the original
EPR states with continuous degrees of freedom. In recent years, quantum
nonlocality for position-momentum variables associated with the original EPR
states was analyzed \cite{Bell86,Grangier,Banaszek,exp,Chen}. In particular,
violations of the Bell-type inequalities by the ``regularized'' EPR states
produced in a pulsed nondegenerate optical parametric amplifier was
experimentally observed by using homodyning with weak coherent fields and
photon counting \cite{exp}.

In connection with the applicability of quantum superposition principle on a
macroscopic scale, Schr\"odinger \cite{Schro} described a {\it gedanken}
experiment, in which a cat is placed in a quantum superposition of being
dead and alive while entangled with a single radioactive atom. The
mesoscopic equivalents of the Schr\"odinger-cat states [called hybrid
entangled states (HES) in the subsequent discussion] have been
experimentally realized for a $^9$Be$^{+}$ ion in traps \cite{cat-ion} and
atoms in high-$Q$ cavities \cite{cat-cavity}. Particularly, in the trapped
ion experiment \cite{cat-ion}, the HES were generated by entangling ion's
internal states ($\left| \uparrow ,\downarrow \right\rangle $ in the
terminology of spin-$1/2$ particles) with discrete spectrum and motional
states with continuous spectrum: 
\begin{equation}
\left| {\rm HES}\right\rangle =\frac 1{\sqrt{2}}\left( \left| \uparrow
\right\rangle \left| x_1\right\rangle +\left| \downarrow \right\rangle
\left| x_2\right\rangle \right) ,  \label{escs}
\end{equation}
where the motional states $\left| x_1\right\rangle $ and $\left|
x_2\right\rangle $ of the ion are two distinguishable wave packets of a
harmonic oscillator and thus denote quantum states with continuous
variables. For the atom-cavity system, the entanglement of the type (\ref
{escs}) occurs between a microwave cavity field and an atom \cite{cat-cavity}%
. These HES are of great theoretical interest in addressing some fundamental
issues, such as decoherence and the quantum/classical boundary \cite
{cat-ion,cat-cavity,Haroche}. The trapped-ion system is a strong candidate
for quantum computation \cite{QIT,Cirac-Z,ion-FP}. In this paper we
demonstrate the HES as a valuable resource in quantum information
processing, building an interesting link between quantum information
protocols of discrete and continuous variables. Quantum nonlocality of the
HES is also analyzed by using the recently developed formulation \cite{Chen}.

For usual two-qubit (qubit-$1$ and qubit-$2$) systems, one can introduce the
following Bell-basis spanned by the two-qubit states 
\begin{eqnarray}
\left| \Psi _{1,2}^{\pm }\right\rangle &=&\frac 1{\sqrt{2}}\left( \left|
\uparrow \right\rangle _1\left| \downarrow \right\rangle _2\pm \left|
\downarrow \right\rangle _1\left| \uparrow \right\rangle _2\right) , 
\nonumber \\
\left| \Phi _{1,2}^{\pm }\right\rangle &=&\frac 1{\sqrt{2}}\left( \left|
\uparrow \right\rangle _1\left| \uparrow \right\rangle _2\pm \left|
\downarrow \right\rangle _1\left| \downarrow \right\rangle _2\right) .
\label{basis}
\end{eqnarray}
The pairs of qubits are maximally entangled when they are in these states.
An analogous Bell-basis spanned by four HES 
\begin{eqnarray}
\left| \psi _{1,2}^{\pm }\left( z\right) \right\rangle &=&\frac 1{\sqrt{2}}%
\left( \left| \uparrow \right\rangle _1\left| z\right\rangle _{o2}\pm \left|
\downarrow \right\rangle _1\left| z\right\rangle _{e2}\right) ,  \nonumber \\
\left| \phi _{1,2}^{\pm }\left( z\right) \right\rangle &=&\frac 1{\sqrt{2}}%
\left( \left| \uparrow \right\rangle _1\left| z\right\rangle _{e2}\pm \left|
\downarrow \right\rangle _1\left| z\right\rangle _{o2}\right) ,
\label{cbasis}
\end{eqnarray}
where $\left| z\right\rangle _e$ ($\left| z\right\rangle _o$) is the even
(odd) coherent state defined in terms of the number states $\left|
n\right\rangle $ 
\begin{eqnarray}
\left| z\right\rangle _e &=&(\cosh \left| z\right|
^2)^{-1/2}\sum_{n=0}^\infty \frac{z^{2n}}{\sqrt{(2n)!}}\left|
2n\right\rangle ,  \nonumber \\
\left| z\right\rangle _o &=&(\sinh \left| z\right|
^2)^{-1/2}\sum_{n=0}^\infty \frac{z^{2n+1}}{\sqrt{(2n+1)!}}\left|
2n+1\right\rangle .  \label{eo}
\end{eqnarray}
The HES $\left| \psi _{1,2}^{\pm }\right\rangle $ and $\left| \phi
_{1,2}^{\pm }\right\rangle $ can be created, e.g., by properly tailoring
laser pulses for the trapped-ion system \cite{cat-ion}. A crucial property
of $\left| \psi _{1,2}^{\pm }\right\rangle $ and $\left| \phi _{1,2}^{\pm
}\right\rangle $ is that they have precisely the same amount of the
entanglement entropy (one ebit) as the four Bell-basis states $\left| \Psi
_{1,2}^{\pm }\right\rangle $ and $\left| \Phi _{1,2}^{\pm }\right\rangle $
for any $z$, as can be easily checked. Nevertheless, the HES are generally
not maximally entangled for $z\neq 0$.

To uncover quantum nonlocality of the HES, one needs to consider whether or
not they violate Bell's inequalities \cite{Bell,CHSH,Bell-book}. Here we
show the violation of Bell's inequality due to Clauser, Horne, Shimony and
Holt (CHSH) by the HES \cite{CHSH}. For this purpose, we can introduce the
following ``pseudospin'' operators for a harmonic oscillator \cite{Chen} 
\begin{eqnarray}
s_z &=&\sum_{n=0}^\infty \left[ \left| 2n\right\rangle \left\langle
2n\right| -\left| 2n+1\right\rangle \left\langle 2n+1\right| \right] , 
\nonumber  \label{oe} \\
s_{+} &=&\sum_{n=0}^\infty \left| 2n\right\rangle \left\langle 2n+1\right|
=(s_{-})^{\dagger },  \label{s}
\end{eqnarray}
The operator $s_z$ is the parity operator $(-1)^N$ ($N$ is the number
operator), while $s_{+}$ and $s_{-}$ are the ``parity-flip'' operators. Then
the commutation relations 
\begin{equation}
\left[ s_z,s_{\pm }\right] =\pm 2s_{\pm },\;\;\;\;\left[ s_{+},s_{-}\right]
=s_z  \label{comm}
\end{equation}
immediately follow from Eq. (\ref{s}) and are identical to those of the spin-%
$1/2$ systems. Therefore pseudospin operators ${\bf \hat s}=(s_x,s_y,s_z)$,
where $s_x\pm is_y=2s_{\pm }$, have the same algebra as the spin operator $%
{\bf \hat \sigma }$. Let us define the following operator (the ``Bell
operator'' \cite{Bell-op}) 
\begin{eqnarray}
{\cal B} &=&({\bf a}\cdot {\bf \hat \sigma })\otimes ({\bf b}\cdot {\bf \hat 
s})+({\bf a}\cdot {\bf \hat \sigma })\otimes ({\bf b}^{\prime }\cdot {\bf 
\hat s})  \nonumber \\
&&+({\bf a}^{\prime }\cdot {\bf \hat \sigma })\otimes ({\bf b}\cdot {\bf 
\hat s})-({\bf a}^{\prime }\cdot {\bf \hat \sigma })\otimes ({\bf b}^{\prime
}\cdot {\bf \hat s}).  \label{bellop}
\end{eqnarray}
Here ${\bf a}$, ${\bf a}^{\prime }$, ${\bf b}$ and ${\bf b}^{\prime }$ are
four unit vectors, e.g., ${\bf a}=(\sin \theta _a,0,\cos \theta _a)$ with $%
\theta _a$ being the ``polar'' angle of ${\bf a}$. Obviously, $({\bf a}\cdot 
{\bf \hat \sigma })^2=({\bf a}^{\prime }\cdot {\bf \hat \sigma }%
)^2=I_{2\times 2}$ and $({\bf b}\cdot {\bf \hat s})^2=({\bf b}^{\prime
}\cdot {\bf \hat s})^2=I$, where $I_{2\times 2}$ ($I$) is the identity
operator in the Hilbert space of the discrete variable (continuous variable)
states. This fact implies that eigenvalues of ${\bf a}\cdot {\bf \hat \sigma 
}$ and ${\bf b}\cdot {\bf \hat s}$ are $\pm 1$, similarly to the usual spin-$%
1/2$ systems. Then local realistic theories impose the following Bell-CHSH
inequality \cite{CHSH}: 
\begin{equation}
\left| \left\langle {\cal B}\right\rangle \right| \leq 2,  \label{chsh}
\end{equation}
where $\left\langle {\cal B}\right\rangle $ is the expectation value of $%
{\cal B}$ with respect to a given quantum state of the present
Schr\"odinger-cat-like system. Quantum mechanically, $\left| \left\langle 
{\cal B}\right\rangle \right| $ is bounded by $2\sqrt{2}$, known as the
Cirel'son bound \cite{Bell-op,MAX}.

Now we can calculate $\left\langle {\cal B}\right\rangle $ with respect to $%
\left| \psi _{1,2}^{\pm }\right\rangle $ and $\left| \phi _{1,2}^{\pm
}\right\rangle $. For example, 
\begin{eqnarray}
\left\langle \phi _{1,2}^{+}\right| {\cal B}\left| \phi
_{1,2}^{+}\right\rangle &=&\cos \theta _a\cos \theta _b+K\sin \theta _a\sin
\theta _b  \nonumber \\
&&\ +\cos \theta _a\cos \theta _{b^{\prime }}+K\sin \theta _a\sin \theta
_{b^{\prime }}  \nonumber \\
&&\ +\cos \theta _{a^{\prime }}\cos \theta _b+K\sin \theta _{a^{\prime
}}\sin \theta _b  \nonumber \\
&&\ -\cos \theta _{a^{\prime }}\cos \theta _{b^{\prime }}-K\sin \theta
_{a^{\prime }}\sin \theta _{b^{\prime }},  \label{bchsh}
\end{eqnarray}
where 
\begin{eqnarray}
K(z) &\equiv &\left. _{2e}\left\langle z\right| s_{2+}\left| z\right\rangle
_{o2}\right.  \nonumber  \label{k} \\
\ &=&(\frac 12\sinh 2z^2)^{-1/2}\sum_{n=0}^\infty \frac{z^{4n+1}}{\sqrt{%
(2n)!(2n+1)!}}<1,  \label{k}
\end{eqnarray}
and we have chosen $z$ to be positive without loss of generality. Setting $%
\theta _a=0$, $\theta _{a^{\prime }}=\pi /2$ and $\theta _b=\tan
^{-1}K=-\theta _{b^{\prime }}$, we have 
\begin{equation}
\left\langle \phi _{1,2}^{+}\right| {\cal B}\left| \phi
_{1,2}^{+}\right\rangle =2\sqrt{1+K^2}.  \label{max}
\end{equation}
Similar results can be obtained for $\left| \phi _{1,2}^{-}\right\rangle $
and $\left| \psi _{1,2}^{\pm }\right\rangle $, indicating that the HES
always violate the Bell-CHSH inequality (\ref{chsh}). However the violation,
which depends on the overlap between $s_{2+}\left| z\right\rangle _{o2}$ and 
$\left| z\right\rangle _{e2}$, is not maximal ($K<1$) unless $z=0$ [$%
K(z=0)=1 $]. It is interesting to compare our result with Ref. \cite
{qp-0010062}, where a similar problem has been considered in a different
route.

Now some remarks are in order. The above discussion on quantum nonlocality
of the HES is applicable when the two parties involved in the HES are
space-like separated. For the atom-cavity system, the requirement of the
space-like separation between a microwave cavity field and an atom can be
easily imposed. However, this is not the case for the trapped-ion system,
where entanglement occurs between two different degrees of freedom of a
single ion and as such, the space-like separation can not be achieved. Thus
for the latter system, it will be more appropriate to consider quantum
contextuality \cite{KS,Peres-book,Michler}. Non-contextual hidden variable
theories predict that the value of an observable is predetermined and thus
independent on the experimental context, i.e., what other co-measurable
observable is simultaneously measured, and whether or not the space-like
separation condition is fulfilled. In fact, the HES for the trapped-ion
system might be an alternative single-particle state that is suitable for
testing the non-contextual hidden variable theories versus quantum
mechanics. Other single-particle states for this purpose have been proposed
in Ref. \cite{Michler}. There is also an issue on how to measure the
pseudospin operators ${\bf \hat s}$ experimentally. In Ref. \cite{Chen}, a
generic, yet feasible, approach has been suggested for achieving this,
though it is experimentally challenging.

As is now well known, the nonlocal correlations, as uncovered here for the
HES, can be exploited to perform classically impossible tasks in the context
of quantum information theory. But for the trapped-ion realization of the
HES, quantum contextuality, rather than quantum nonlocality, might also be
of practical importance for the quantum information tasks, as will become
clear later.

Quantum teleportation is a process that transmits an unknown quantum state
from a sender (Alice) to a receiver (Bob) via a quantum channel with the
help of some classical information. For transmitting the unknown qubit state
with fidelity $1$, the quantum channel is a maximally entangled state (e.g., 
$\left| \Psi _{1,2}^{+}\right\rangle $) \cite{telep-d}. It can be the HES
realized with either the atom-cavity system or the trapped-ion one in the
present case. Here we propose a protocol using $\left| \phi
_{2,3}^{+}\right\rangle $ (one of the HES of the trapped-ion system) hold by
Alice as the quantum channel to ``teleport'' an unknown spin state of
another ion (initially hold also by Alice) 
\begin{equation}
\left| \varphi \right\rangle _1=\alpha \left| \uparrow \right\rangle
_1+\beta \left| \downarrow \right\rangle _1  \label{send}
\end{equation}
with $\left| \alpha \right| ^2+\left| \beta \right| ^2=1$. The initial state
of the whole system before teleportation is therefore 
\begin{eqnarray}
\left| \varphi \right\rangle _1\left| \phi _{2,3}^{+}\right\rangle &=&\frac 1%
2[\left| \Phi _{1,2}^{+}\right\rangle (\alpha \left| z\right\rangle
_{e3}+\beta \left| z\right\rangle _{o3})  \nonumber \\
&&+\left| \Phi _{1,2}^{-}\right\rangle (\alpha \left| z\right\rangle
_{e3}-\beta \left| z\right\rangle _{o3})  \nonumber \\
&&+\left| \Psi _{1,2}^{+}\right\rangle (\alpha \left| z\right\rangle
_{o3}+\beta \left| z\right\rangle _{e3})  \nonumber \\
&&+\left| \Psi _{1,2}^{-}\right\rangle (\alpha \left| z\right\rangle
_{o3}-\beta \left| z\right\rangle _{e3})].  \label{bef}
\end{eqnarray}

Now, similarly to the original proposal \cite{telep-d}, Alice performs the
spin Bell-state measurement with four measurement outcomes ($\Phi
_{1,2}^{\pm }$ and $\Phi _{1,2}^{\pm }$, each with probability $1/4$).
According to the standard quantum measurement theory, after her measurement,
the motional state in the quantum channel must be one of the following four
states 
\begin{eqnarray}
&&\ \alpha \left| z\right\rangle _{e3}+\beta \left| z\right\rangle
_{o3},\;\;\alpha \left| z\right\rangle _{e3}-\beta \left| z\right\rangle
_{o3},  \nonumber \\
&&\ \alpha \left| z\right\rangle _{o3}+\beta \left| z\right\rangle
_{e3},\;\;\alpha \left| z\right\rangle _{o3}-\beta \left| z\right\rangle
_{e3}.  \label{four}
\end{eqnarray}
In the case of the first outcome $\left| \Phi _{1,2}^{+}\right\rangle $, the
state $\alpha \left| z\right\rangle _{e3}+\beta \left| z\right\rangle _{o3}$
has already been a replica of $\left| \varphi \right\rangle _1$; but now the
``parity state'' $\left| z\right\rangle _{e3}$ ($\left| z\right\rangle _{o3}$%
) with parity $+1$ ($-1$) plays the same role as the spin state $\left|
\uparrow \right\rangle _3$ ($\left| \downarrow \right\rangle _3$). In the
remaining three cases, Alice needs to perform one of the unitary operations $%
(s_{3z},s_{3x},s_{3y})$, yielding, respectively, $\alpha \left|
z\right\rangle _{e3}+\beta \left| z\right\rangle _{o3}$, $\alpha
(s_{3+}\left| z\right\rangle _{o3})+\beta (s_{3-}\left| z\right\rangle
_{e3}) $, and $-i[\alpha (s_{3+}\left| z\right\rangle _{o3})+\beta
(s_{3-}\left| z\right\rangle _{e3})]$. In this way Alice's motional state in
the quantum channel is converted into a replica of her spin state $\left|
\varphi \right\rangle _1$ (except for an irrelevant phase factor). Note here
that $s_{3+}\left| z\right\rangle _{o3}$ ($s_{3-}\left| z\right\rangle _{e3}$%
) has parity $+1$ ($-1$) and thus plays again the same role as the spin
state $\left| \uparrow \right\rangle _3$ ($\left| \downarrow \right\rangle
_3 $). The remaining three HES in Eq. (\ref{cbasis}) can also be used as the
quantum channel.

The present teleportation protocol can be regarded as a realization of
continuous variable qubit encoded in parity: Though the teleported state $%
\alpha \left| z\right\rangle _{e3}+\beta \left| z\right\rangle _{o3}$ has
continuous spectrum, it carries the same information as a usual qubit when
only parity measurement is performed. Such a qubit encoding for a single
bosonic mode has been proposed in Ref. \cite{Milburn}. It is also
interesting to compare the present teleportation scheme to the
``two-particle scheme'' for quantum teleportation \cite{Italy} when the HES
of the trapped-ion system are used: Here the quantum channel consists of
entanglement between two different degrees of freedom of a single ion.

Similarly to teleporting the spin state (\ref{send}), Alice can also
teleport an unknown state $\left| \varphi (z^{\prime \prime })\right\rangle
_3=\alpha \left| z^{\prime \prime }\right\rangle _{e3}+\beta \left|
z^{\prime \prime }\right\rangle _{o3}$ of a continuous variable qubit via
one of $\left| \psi _{1,2}^{\pm }\right\rangle $ and $\left| \phi
_{1,2}^{\pm }\right\rangle $, converting her state into (\ref{send}). In
this case, Alice needs to perform the ``parity Bell-state measurement''
collapsing her state into one of the four ``parity Bell-basis'' states: 
\begin{eqnarray}
\left| \tilde \phi _{1,2}^{\pm }\left( z,z^{\prime }\right) \right\rangle &=&%
\frac 1{\sqrt{2}}\left( \left| z\right\rangle _{e1}\left| z^{\prime
}\right\rangle _{e2}\pm \left| z\right\rangle _{o1}\left| z^{\prime
}\right\rangle _{o2}\right) ,  \nonumber \\
\left| \tilde \psi _{1,2}^{\pm }\left( z,z^{\prime }\right) \right\rangle &=&%
\frac 1{\sqrt{2}}\left( \left| z\right\rangle _{e1}\left| z^{\prime
}\right\rangle _{o2}\pm \left| z\right\rangle _{o1}\left| z^{\prime
}\right\rangle _{e2}\right) .  \label{zz}
\end{eqnarray}
Any of $\left| \tilde \psi _{1,2}^{\pm }\left( z,z^{\prime }\right)
\right\rangle $ and $\left| \tilde \phi _{1,2}^{\pm }\left( z,z^{\prime
}\right) \right\rangle $ has one ebit of entanglement for any $z$ and $%
z^{\prime }$ and thus can be used to teleport one parity qubit. The parity
Bell states (\ref{zz}) can be regarded as ``entangled two-cat states''
consisting of two macroscopically distinguishable wave packets. A related
issue in this respect is the entangled coherent states \cite{ent-cs}.

Entanglement swapping is in fact teleportation of entanglement \cite
{telep-d,swap-d}. Here we consider entanglement swapping between two HES $%
\left| \phi _{1,2}^{-}\left( z\right) \right\rangle $ and $\left| \phi
_{3,4}^{-}\left( z^{\prime }\right) \right\rangle $. In terms of the spin
Bell-basis (\ref{basis}) and the parity Bell-basis (\ref{zz}), the initial
state before entanglement swapping is 
\begin{eqnarray}
\left| \chi \right\rangle &=&\left| \psi _{1,2}^{-}\left( z\right)
\right\rangle \left| \psi _{3,4}^{-}\left( z^{\prime }\right) \right\rangle 
\nonumber \\
\ &=&\frac 12\left[ \left| \Phi _{1,3}^{+}\right\rangle \left| \tilde \phi
_{2,4}^{+}\left( z,z^{\prime }\right) \right\rangle -\left| \Phi
_{1,3}^{-}\right\rangle \left| \tilde \phi _{2,4}^{-}\left( z,z^{\prime
}\right) \right\rangle \right.  \nonumber \\
&&\ -\left. \left| \Psi _{1,3}^{+}\right\rangle \left| \tilde \psi
_{2,4}^{+}\left( z,z^{\prime }\right) \right\rangle +\left| \Psi
_{1,3}^{-}\right\rangle \left| \tilde \psi _{2,4}^{-}\left( z,z^{\prime
}\right) \right\rangle \right] .  \label{swap}
\end{eqnarray}
By performing a joint spin Bell-state measurement, the discrete-variable
state is projected onto one of the Bell states (\ref{basis}). This
measurement automatically collapses the continuous-variable states into one
of the parity Bell state (\ref{zz}). This entanglement swapping procedure
can thus be regarded as producing the parity Bell state (\ref{zz}).

When the quantum channel is the HES\ of the atom-cavity system, the above
protocol is still valid. But an crucial difference is that now both Alice
and Bob are involved in the teleportation, as in the original quantum
teleportation protocol. Thus, the quantum state transfer may be implemented
with two distinct quantum channels. For the HES of the atom-cavity system,
the quantum channel possesses nonlocal correlations that are essential for
succeeding in quantum teleportation. However, for the trapped-ion
realization of the HES, there is no quantum nonlocality, but quantum
contextuality; in this case quantum ``teleportation'' is in fact the quantum
state local (not remote) transfer. Actually, in the context of quantum
information processing it is more important to consider the utility of
quantum correlations \cite{Vaidman}, e.g., quantum nonlocality and quantum
contextuality.

How practical are the present protocols on quantum teleportation and
entanglement swapping? Here we consider this problem by taking the
trapped-ion system as an example. For the purpose of high precision
spectroscopy and frequency standard, preparation and manipulation of quantum
states of the trapped ion system are a mature technology. The internal
states of ions can be measured using the quantum jump technique with nearly $%
100\%$ detection efficiency \cite{QIT}. Recently, it has been shown \cite
{Wineland-Solano} that we are able to deterministically generate all the
spin Bell states (\ref{basis}), which can then be detected by resonance
fluorescence shelving methods. Thus the present protocol of entanglement
swapping should be realizable with current experimental technology. To
implement perfect teleportation of the spin state (\ref{send}) via the
shared entanglement $\left| \phi _{2,3}^{+}\right\rangle $, one needs to
perform the local unitary operations $(s_{3z},s_{3x},s_{3y})$. But
presently, how to practically realize these operations still remains an open
question. Nevertheless, the teleportation protocol does succeed in
teleporting (\ref{send}) into the parity state $\alpha \left| z\right\rangle
_{e3}+\beta \left| z\right\rangle _{o3}$ faithfully, or up to one of the
local operations $(s_{3z},s_{3x},s_{3y})$. In the latter three cases, the
output states are the transformed version of the desired parity qubits.
Similarly to the analysis made in Ref. \cite{tele-U}, this feature of the
protocol might be potentially useful, e.g., in realizing the
difficult-to-implement logic operations $(s_z,s_x,s_y)$.

The ions in traps are a promising system to implement quantum computing \cite
{QIT,Cirac-Z}. In such a Cirac-Zoller quantum computer, the relevant
motional states of ions are $\left| 0\right\rangle $ and $\left|
1\right\rangle $ only, approximately treating the harmonic oscillator as a
two-level system. The teleportation protocol proposed in this work has
realized the continuous variable parity qubits. This might motivate quantum
computing with these parity qubits. Several essential components for this
purpose have already been demonstrated in Ref. \cite{Milburn}. These include
the unitary construction of parity qubits, one and two qubit logical
operations on the parity states and error correction of the qubits. The
two-qubit (one is the usual qubit and the other is the parity qubit) logical
operations is also possible \cite{Milburn}. Thus it seems feasible that the
trapped-ion system can be used to implement quantum computation on the
hybrid (parity and spin) qubits.

Finally, we point out that the parity qubits, including these in quantum
channels, are encoded by odd and even coherent states in our discussion.
However there is, in principle, no reason to insist on such an encoding. In
fact, one can also encode the parity qubits as $\left| {\rm p}\right\rangle
=\alpha \left| +\right\rangle +\beta \left| -\right\rangle $, where $\left|
+\right\rangle $ ($\left| -\right\rangle $) denotes an arbitrary parity
state with parity $+1$ ($-1$). The complex probability amplitudes $\alpha $
and $\beta $ represent quantum information when only the parity measurement
is concerned; the remaining unknown quantum information in $\left|
+\right\rangle $ and $\left| -\right\rangle $ is meaningful merely in
continuous variable quantum protocols. When applied to the atom-cavity
system, the proposed scheme might be useful in quantum information
processing based on the quantum network consisting of many atom-cavities
(``nodes'') connected by light field \cite{network}, whose parity states
carry quantum information.

In summary, the HES have been shown to violate the Bell-CHSH inequality.
Quantum nonlocality of the HES is thus uncovered. We then demonstrate the
HES, each of which carries exactly one ebit of entanglement, as a valuable
resource in quantum information processing, such as quantum teleportation,
entanglement swapping and quantum computation over continuous variable
parity qubits. Since the entanglement of the HES occurs between the states
with both discrete spectrum and with continuous spectrum, our work
establishes an interesting link between quantum information protocols of
discrete and continuous variables.

This work was supported by the National Natural Science Foundation of China
under Grants No. 10104014, No. 19975043 and No. 10028406, and by the Chinese
Academy of Sciences.

\end{document}